\documentclass{sig-alternate}

\usepackage[caption=false,font=footnotesize]{subfig}
\graphicspath{{./fig/}}
\DeclareGraphicsExtensions{.pdf,.jpeg,.jpg,.png}
\usepackage{hyperref}
\usepackage{epsfig,endnotes}
\usepackage{graphicx,calc,float}
\usepackage{endnotes,microtype,xspace,fancyvrb,multirow}
\usepackage{array,underscore,relsize}
\usepackage{booktabs}
\usepackage{tabularx}
\usepackage{xcolor}
\usepackage{pifont}

\newcommand{\cmark}{\ding{51}}%
\newcommand{\xmark}{\ding{55}}%

\begin{document}
%
\title{Cloaker Catcher:\\
	A Client-based Cloaking Detection System}

%
%
%
%
%

\numberofauthors{3} 
%
\author{
%
%
\alignauthor
Ruian Duan\\
       \affaddr{Georgia Institute of Technology}\\
       \email{ruian@gatech.edu}
\alignauthor
Weiren Wang\\
       \affaddr{Georgia Institute of Technology}\\
       \email{weirenwang@gatech.edu}
\alignauthor Wenke Lee\\
       \affaddr{Georgia Institute of Technology}\\
       \email{wenke.lee@gmail.com}
}

\maketitle

\begin{abstract}
Cloaking has long been exploited by spammers for the purpose of increasing the exposure of their websites. In other words, cloaking has long served as a major malicious technique in search engine optimization (SEO). Cloaking hides the true nature of a website by delivering blatantly different content to users versus web crawlers. Recently, we have also witnessed a rising trend of employing cloaking in search engine marketing (SEM). However, detecting cloaking is challenging. Existing approaches cannot detect IP cloaking and are not suitable for detecting cloaking in SEM because their search-and-visit method leads to click fraud. In addition, they focus on detecting and measuring cloaking on the server side, but the results are not visible to users to help them avoid frauds.

Our work focuses on mitigating IP cloaking and SEM cloaking, and providing client-based real-time cloaking detection services. To achieve these goals, we first propose the Simhash-based Website Model (SWM), a condensed representation of websites, which can model natural page dynamics. Based on SWM, we design and implement Cloaker Catcher, an accurate, efficient and privacy-preserving system, that consists of a server that crawls websites visited by users on demand and a client-side extension that fetches spider views of websites from the server and compares them with user views to detect cloaking. Since Cloaker Catcher checks on the client side for each real user, IP cloaking can be detected whenever it occurs and click fraud in SEM can also be prevented. Using our system, we conducted the first analysis of SEM cloaking and found that the main purpose of SEM cloakers is to provide illicit services.

\end{abstract}

\terms{Security, Privacy, Spam}


\section{Introduction}
\label{s:intro}
Today, popular search engines such as Google, Bing and Yahoo, are the main entries to the information on the Internet for users. By their abilities to drive web traffic and connect users with retailers, search engines have changed the advertising history. Search engines monetize their traffic either via ``organic" search results or sponsored search placements - together comprising a \$24B marketing sector~\cite{marketsize}.

The great value of search engine results has attracted public attention, and two major approaches have been applied to increase website visibility through search engines. The first approach, Search Engine Optimization (SEO) is the process of affecting the visibility of a website or a web page in a search engine's unpaid results. The second, Search engine marketing (SEM),
\footnote{While search engine marketing can be a broader concept about marketing related to search engines, this paper refers to SEM as search advertisements, i.e. sponsored ads on search engines}
refers to the promotion of websites by increasing their visibility through sponsored search placements, i.e. search ads.
Search engine operators encourage the use of certain techniques such as optimizing content in SEO or selecting long-tail advertising keywords in SEM. However, search engines explicitly ban techniques, such as link farms, keyword stuffing and cloaking, which are specifically designed to cheat and manipulate rankings.
Among these techniques, cloaking is the basic step for serving non-compliant content.
It has been widely employed to hide the true nature of websites, because of its low setup cost and the lack of effective and efficient detection methods used against it.
While search operators try to identify and remove pages that host harmful content (e.g. phishing pages, malware sites, etc.), scammers seek to elude such detection using cloaking.
Typically, a cloaker serves ``benign" content to crawlers and scam content to normal users who are referred via a particular search request.
An SEM cloaking example is that, a cloaker can advertise illegal goods or services (e.g. drugs, ghostwriting service) on sponsored ads, but claim themselves as innocent ads to search engines.
Serving such illicit advertisements not only put consumers at risk, but also cause high cost to search operators, e.g. Google paid \$500 million to settle a lawsuit with the U.S. Department of Justice for accepting advertisements from unlicensed pharmacies~\cite{adsfine}.

\textbf{Technical Challenges.}
Various approaches have been proposed to detect cloaking sites and measure their prevalence. Previous detection studies~\cite{wu2006detecting, wang2006detecting, chellapilla2006improving, lin2009detection, deng2013uncovering, wang2011cloak, najork2003system} focused mainly on using heuristics such as search query monetizability, features in websites such as page size and HTML tags, to improve detection accuracy.
However, these approaches have three drawbacks. First, they cannot handle IP cloaking because they fail to use search engine IPs to fetch spider views and real user IPs to fetch user views.
Second, they are designed for SEO cloaking detection and are not suitable for SEM cloaking because these methods require inspectors to pretend to be real users and click on search results, but doing so in SEM leads to click fraud and breaks the business model of SEM. Third, their cloaking decisions are made on the server side and cannot directly protect users.

\textbf{Our Solution.}
Instead of leaving all the workload and decision making to servers, we propose Cloaker Catcher, a client-based system that detects cloaking on the user side in real-time. Our system overcomes the three drawbacks shared by the previous systems. First, since detection is done on the user side,
the system obtains valid user IPs. Using Google Translate, the system gets search engine IPs. Therefore, cloakers have no place to hide and IP cloaking will be detected whenever they occur. Second, since cloaking decision is made for each user without click fraud, detecting cloaking in SEM is therefore the same as in SEO. Third, our system detects cloaking when user clicks the search results or ads and provides real-time responses to users. Furthermore, this cloaking signal can be combined with other features such as DNS domains~\cite{lu2011surf} to improve accuracy or integrated into existing APIs such as Safe Browsing API ~\cite{safebrowsing} to guard users' browsing sessions.

Apart from differentiating cloaking sites from dynamic sites, a client-based system has two additional requirements: (a) since user machines are resource constrained, the system should introduce low overhead, including both computation and network traffic,
(b) the system should not have access to the users' copy of websites, because what user sees is highly private. In short, we need a cloaking detection system that is accurate, efficient, privacy-preserving, and minimizes data transmission.
To meet these requirements, this paper proposes a new way to model website dynamics based on a locality sensitive hash algorithm, Simhash. We build fuzzy signatures of contents and the layout for websites. We then learn patterns, namely, Simhash Website Models (SWM), for each website and then use them to model the dynamics of websites.
SWM has three significant advantages. They are compact, easy to compare and accurate in cloaking detection. We build Cloaker Catcher based on SWM and achieve a 97.1\% true positive rate at only a 0.3\% false positive rate with low user overhead.

With Cloaker Catcher, we further conduct a measurement to understand current state of SEO and SEM cloaking on Google search. Our results show that cloaking is more popular in SEO than SEM and the primary goal in SEM cloaking is to provide illegal service, in contrast to traffic sale in SEO.

\textbf{Roadmap.} The remainder of this paper is structured as follows.  ~\autoref{s:overview} provides the necessary background and challenges, and proposes our solution. ~\autoref{s:design} introduces Simhash-based Website Models (SWM) and a client-based detection system based on SWM. ~\autoref{s:impl} describes our data collection process, model training and configuration of Cloaker Catcher. ~\autoref{s:measurement} measures the current cloaking state in SEO and SEM, followed by performance evaluation of our system and comparison against previous works in ~\autoref{s:performance}. ~\autoref{s:limitation} discusses limitations of our approach.
~\autoref{s:relwk} describes related work on search engine spams and simhash. ~\autoref{s:conclusion} concludes this work.

\section{Overview}
\label{s:overview}
Search engines have been the main sources of information for users since early 21st century. By typing keywords in search engines, users can get desired information in seconds. The ability of search engines to direct web traffic created strong incentives to manipulate the results of Page Rank algorithms and promote pages in search result listings, and have attracted wide attention, from both online market and researchers.
Online marketers use various techniques to promote their ranking in relevant searches, which leads to higher daily visits and profits. Scammers do the same, except that they use ``blackhat" techniques and try to fool users and search operators. Researchers and search operators identify these scammers based on features and penalize their sites. 
However, scammers use cloaking as a standard tool to obscure search engines' inspection and add significant complexity for them to differentiate legitimate web content from fraudulent pages.
Our work tackles the cloaking problem and reveal activities of these scammers.

\subsection{Background and Example}
\label{ss:overview1}
\textbf{Cloaking Types.}
The key technique of cloaking is to distinguish users and crawlers.
Depending on how users and crawlers are identified, cloaking techniques are classified as repeat cloaking, user-agent cloaking, referer cloaking, Javascript Cloaking, and IP cloaking.
In repeat cloaking, victims are tracked on the user-side via cookies or the server-side
via server log. User-agent cloaking refers to websites that check the user-agent string in HTTP request to identify crawlers. Similarly, referer cloaking is done by checking the referer field to identify users redirected from search engine sites. Javascript cloaking works by fingerprinting browsers and serving abusive payload only to real users, rather than spiders.
In IP cloaking, scammers maintain a list of known crawler IPs or inspectors, such as security companies, and serve benign content to these them. IP cloaking requires more work, but is more effective. Crawler IPs are easily available online and some commercial tools, such as noIPfraud and wpCloaker~\cite{intro:cloakware}, even periodically update their lists. In practice, different types of cloaking are usually used in combination, making them very hard to detect.

\textbf{Cloaking Example.}
To concrete the idea of cloaking, we present one example, a plagiarism service in sponsored ads. In ~\autoref{sem:search}, a cloaking ad is placed for the term ``Essay Writing". The cloaker provides users plagiarism service which is against Google's ad policy~\cite{google-ad-policy}, and remains invisible to search operators. In the background, this website use the user-agent and referer to distinguish visitors and hide from inspectors. 

\subsection{Threat Model}
\label{ss:overview2}
\begin{figure}[h]
	\centering
	\includegraphics[width=.45\textwidth]{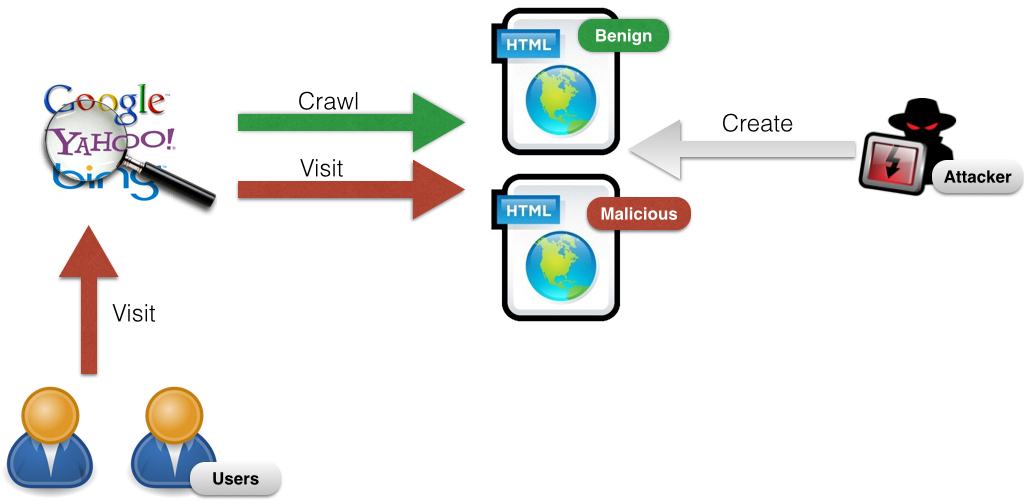}
	\vspace{-1em}
	\caption{Thread Model}
	\vspace{-1.5em}
	\label{fig:threat-model}
\end{figure}
There are multiple entities involved in SEO and SEM cloaking, including users, search engines and website owners (advertisers in SEM). Search engines periodically crawl and index websites and ensure quality of service by considering various kinds of features in their ranking algorithm. Site owners want to drive traffic to their sites by ranking high in search results or ads. Users query terms on search engines and click on the returned links.
As shown in ~\autoref{fig:threat-model}, in cloaking detection, site owners are the attackers and fool both search engines and users. The attackers craft two versions of their websites, and send the benign version to search engines for indexing and ranking, and the malicious version to users for profit. The spiders cannot see the scam content, otherwise, search engines or researchers could use other tools to analyze the page, understand what is being displayed and flag appropriately.
Therefore, in this paper we trust what spiders see and distrust what users see, and compare spider views with user views to find attackers.

\subsection{Client-based Cloaking Detection}
\label{ss:overview3}
Although many systems~\cite{najork2003system, wu2006detecting, wang2006detecting, chellapilla2006improving, lin2009detection, deng2013uncovering, wang2011cloak} have been proposed to detect cloaking, they share three drawbacks. First, they cannot detect IP cloaking. This is because these systems use centralized servers to collect user views, while scammers and commercial cloaking tools can identify their servers and send crafted benign content instead of the illicit ones. Second, most of them are designed for detecting cloaking in SEO and are not suitable for SEM.
The reason is that these methods require inspectors to pretend to be real users and click on search results, and the click-through method in SEM leads to click fraud and breaks the business model of SEM, where one Pay-Per-Click ad can cost more than 100\$.
One exception is Najork's system~\cite{najork2003system} which uses a toolbar to send the signature of user perceived pages to search engines. But they cannot handle dynamic websites, and raises high false positive.
Third, these systems detect cloaking on the server side and cannot protect individual user visits in real-time. Wang et al~\cite{wang2011cloak} measured the search engine response time to cloaking practice and the cloaking duration of websites. The search engine response time is defined as the time from cloaking pages show up in top 100 pages for a specific keyword until they are not. The cloaking duration refers to how long until cloaked pages are no longer cloaked. Google took around two days to remove more than 50\% of the cloaked results for hot search words, and one week for pharmaceutical words, which are abuse-oriented. Yahoo took longer. In terms of cloaking duration, over 80\% of these pages remained cloaked after 7 days.
This means many users have visited those websites before search engines act on cloaking sites. In addition, users may revisit those sites, because they are previously returned as ``trustworthy" results.

Instead of leaving all the work to search engines, we want to leverage the fact that cloakers need to reach end users and catch cloakers on the user side. There are three challenges in client-based cloaking detection. First and foremost, the algorithm should be accurate despite presence of dynamic websites. Second, the system should introduce low overhead to users, including network traffic and  computation. Third, the system should be privacy preserving.
In this paper, we present Cloaker Catcher, to address the three challenges.
Cloaker Catcher consists of a server which crawls websites visited by users on demand and a client-side extension which fetches spider views of websites from the server and compare them with user views to detect cloaking.

\textbf{Accuracy.} Dynamic nature of websites has been the main challenge in cloaking detection.
Previous works introduce different sets of features such as statistical summary, words and HTML tags, and train a fixed threshold for detection. Similarly, this paper employs Simhash~\cite{charikar2002similarity} to generate fuzzy signatures for words and tags on each website and use these signatures to detect cloaking. Simhash is a special signature of a feature set. It is a class of Locality Sensitive Hash (LSH) that has been extensively used in near duplicate detection in search engines.
Simhash is an random projection based algorithm that maps high dimensional data to fixed bits while
maintaining the property that, hamming distance between the resulting bits is an
estimation of the cosine similarity between the original feature set. As the signature is fixed bits (64 in this work), the comparison can be done in $\mathcal{O}(1)$.
We train Simhash-based Website Models (SWM) from multiple spider crawls using hierarchical clustering and compare with user copies to identify outliers, i.e. cloaking candidates. Our approach could achieve 97.1\% TPR and 0.3 FPR. 

\textbf{Efficiency.} Overhead is highly relevant to usability in client-based systems. In our system, the overhead breaks down to three parts: fetch server models, extract client view features and compare the client view with server models. Similar to previous systems, we use words and tags as features, which correspond separately to content and structure of a website. Let $N_W$, $N_T$ be number of words and tags. Server models in our work have constant sizes and introduce low network traffic to users. Feature extraction requires a pass through the website where the complexity is $\mathcal{O}(N_W + N_T)$. Comparison is expensive and computing the greatest common feature subsets between websites requires at least iterating them. Reducing the comparison cost is a well studied problem in near duplicate detection literatures of information retrieval. The standard technique is to use locality sensitive hash to reduce the feature space to fixed bits and compare these compact representations. Inspired by these works, Cloaker Catcher work employs Simhash algorithm to compress feature sets and estimate the distance between user views and spider views.
A comprehensive efficiency measurement is presented in ~\autoref{s:performance}.

\textbf{Privacy.} It is privacy-invasive to disclose users' web views to servers, because these views may contain sensitive data, such as name, email address, credit card number etc. Benefiting from compact model representations, Cloaker Catcher is able to send server views to users and compares on the client side. The visited URLs of users are sensitive as well because they can be used to  identify users. Luckily, Safe Browsing API~\cite{safebrowsing} provides a solution to achieve differential privacy. The basic idea is to receive and send URLs in batches, which prevents servers from inferring the exact visited URL. For simplicity reasons, Cloaker Catcher communication uses the visited link and doesn't implement the URL privacy protection.

\section{System Design}
\label{s:design}
Given the challenges and solutions described in~\autoref{s:overview}, we introduce the Simhash-based Website Model to compactly represent websites and propose Cloaker Catcher based on SWM.

\subsection{Simhash-based Website Model}
\label{ss:swm}
A website is rendered through Document Object Model (DOM), which is maintained by browsers as a tree. A DOM tree contains information about layout of a website (Cascading Style
Sheets is supplemental to DOM in describing the look and
formatting of a document). Among different types of DOM nodes, text nodes represent messages visible to user and are frequently used in cloaking detection literatures. 
This work focuses on structure of DOM tree (tags) and text nodes that user actually sees.
Javascript snippets are not used, since they are mainly used by websites to handle user inputs and manipulate DOM trees, which in turn attracts our attention to the aforementioned features.
In order to model dynamic websites, we propose Simhash-based Website
Models, which use clusters learned from multiple spider copies to quantify average and variance page updates.

\subsubsection{Distance Approximation}
Simhash~\cite{charikar2002similarity} is a dimensionality reduction technique.
It is a fuzzy hash function family that maps a high dimension dataset into fixed
bits and preserves the following properties: 
(A) the fingerprint of a dataset is a ``hash" of its features, and (B) the
hamming distance between two hash values is an estimation of the cosine similarity $\theta(\vec{u}, \vec{v})/\pi$ between the original datasets (represented as vector $\vec{u}, \vec{v}$). This is different from cryptographic hash functions like SHA-1 or MD5, because they will hash two documents which differs by single byte into two completely different hash-values and the hamming distance between them is meaningless.

\subsubsection{Computation}
The computation of Simhash starts from a set of features.
Given a set of features extracted from a document and their corresponding
weights, we use Simhash to generate an f-bit fingerprint as follows.
We maintain an f-dimensional vector V, each of whose dimensions is initialized to zero.
A feature is hashed into an f-bit hash value. These f bits (unique to the
feature) increment/decrement the f components of the vector by the weight of
that feature as follows: if the i-th bit of the hash value is 1, the i-th component
of V is incremented by the weight of that feature; if the i-th bit of the
hash value is 0, the i-th component of V is decremented by the weight of that
feature. When all features have been processed, some components of V are
positive while others are negative. The signs of components determine the
corresponding bits of the final fingerprint. In this work, we set f to 64,
which is the same  effective setting for a corpus of 8 billion websites as described in ~\cite{manku2007detecting}.

There are two characteristics in the computation of the Simhash. First, the order of
the features doesn't matter because Simhash is maintaining a global counter V.
Second, size of the feature set should be relatively large, because Simhash is
random projection based approach, a small set of features may result in completely
different Simhash with one feature difference.
These two characteristics can be indicates that the structural information should be included in the feature set if the order is important and the number of features should be relatively large.

\subsubsection{Feature Extraction}
\begin{figure}[h]
	\centering
	\includegraphics[width=.45\textwidth]{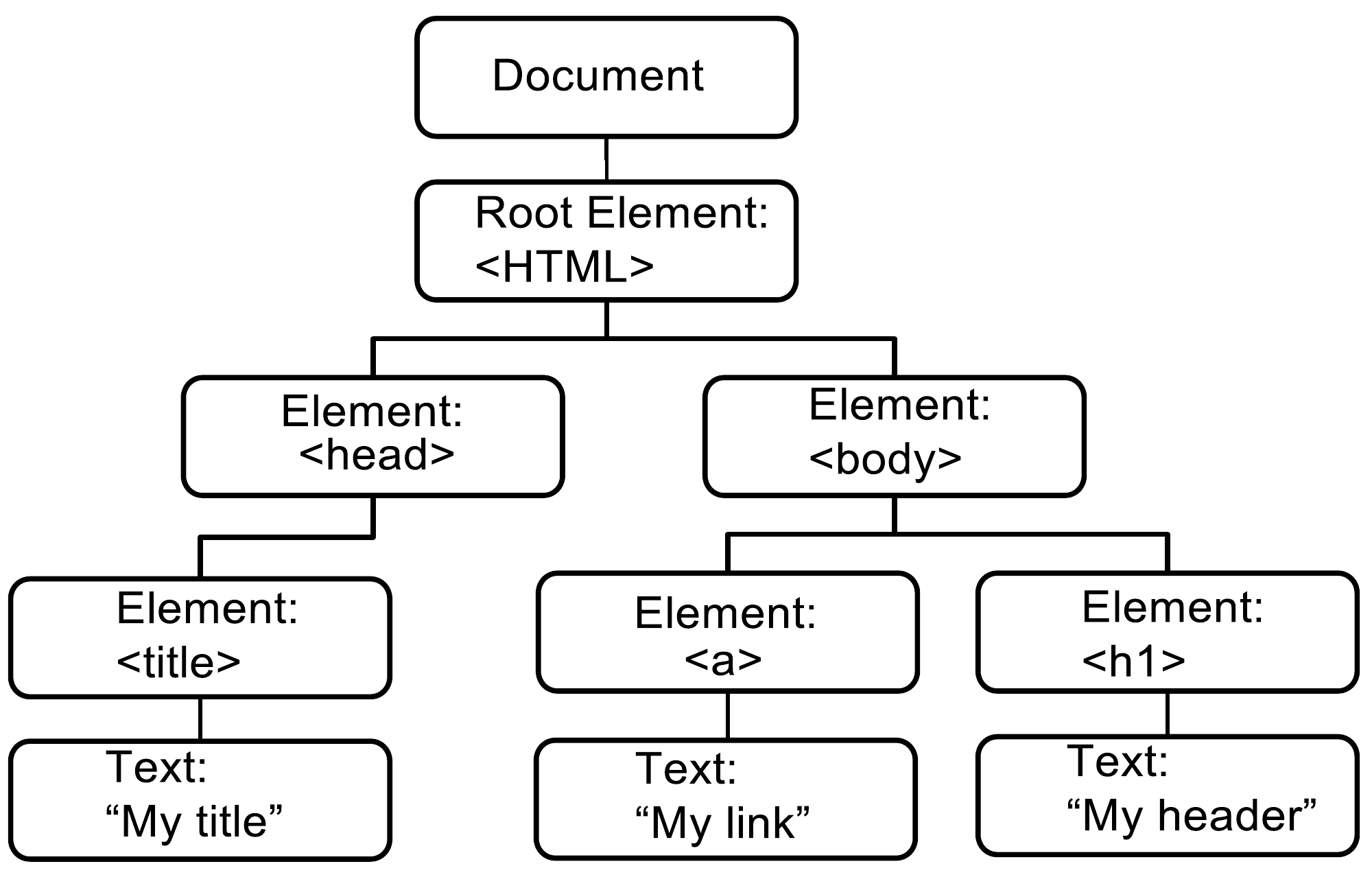}
	\vspace{-1em}
	\caption{DOM tree example}
	\vspace{-1.5em}	
	\label{fig:dom-tree}
\end{figure}

In order to model websites and detect cloaking, we need to capture the behavior and similarity that
a website maintains. We focus on text and tag features, and generate Simhash values separately.
Regarding text features, our algorithm first extracts the visible sequence of words from websites, and then extracts words, bi-gram, tri-gram set (repeated elements only recorded once) from this sequence. For example, the text features of sentence ``i am a cloaker" are \{i, am, a, cloaker, i am, am a, a cloaker, i am a, am a cloaker\}. Because there are usually a large amount of words on a website, and bi-grams and tri-grams represent structure of documents, Simhash is suitable for compressing these features.


Similarly, for tags in a DOM tree, we record presence of each non-text node (tag name), as
well as presence of each child parent pair. The node set records what tag is present in a page, and child parent pair records how these tags are organized, i.e., structure information.
In practice, simply recording the presence of tags yields a relatively small set of features because only a few HTML tags are frequently used. Therefore, we record both the tag name and its associated attribute names to gain more features (higher entropy). Attribute values are discarded because based on our observation, they may change for each visit, especially in SEM. A DOM tree example is shown in ~\autoref{fig:dom-tree}. The corresponding tag features are \{html, head, body, title, \ldots, (head, html), (body, html), \ldots \}.

\begin{figure*}[t]
  \centering
  \subfloat[Text Simhash of Yahoo]{
    \centering
    \includegraphics[width=.37\textwidth]{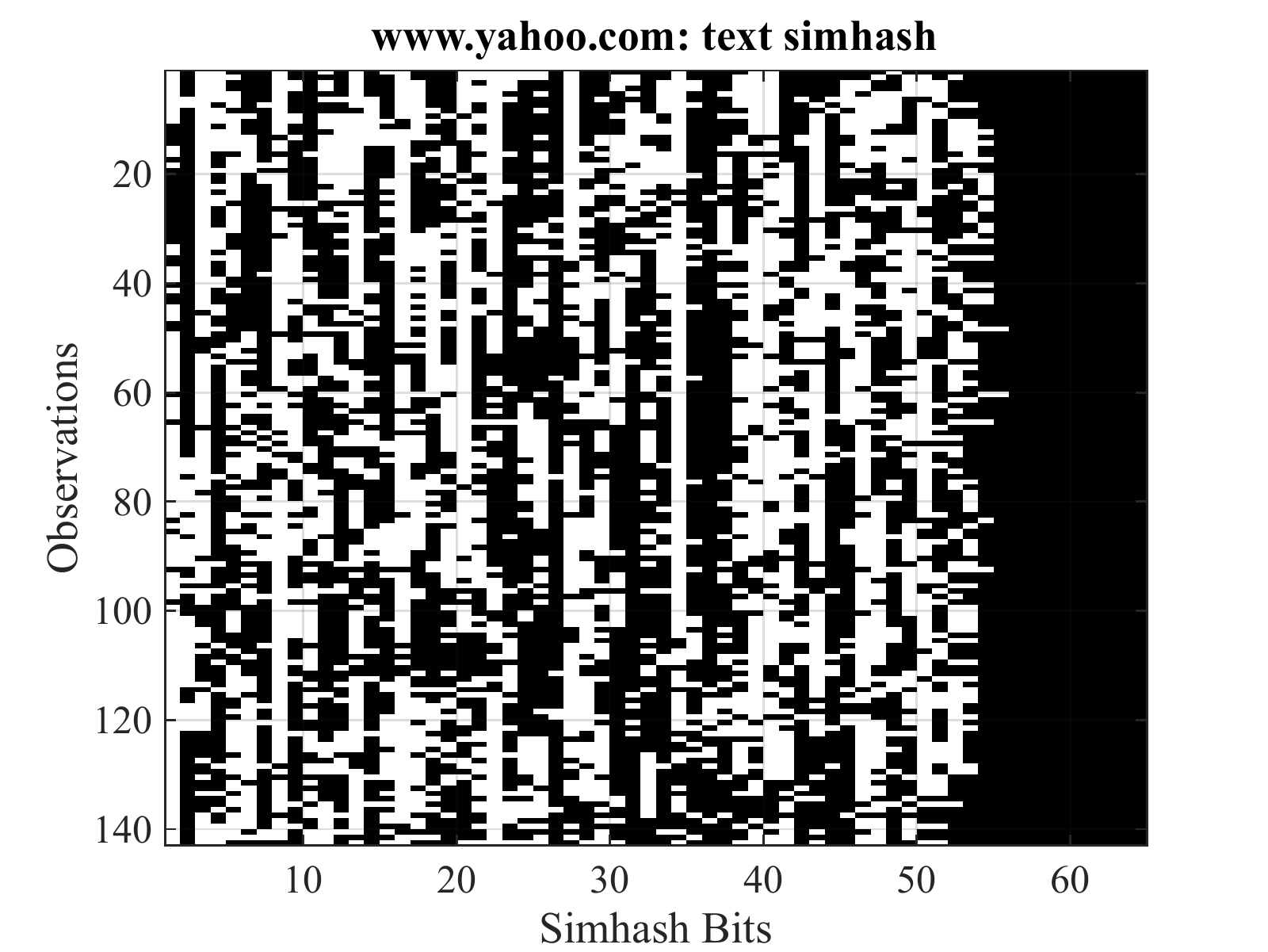}

    \label{fig:yahoo-text-user}
  }
  \subfloat[Tag Simhash of Yahoo]{
    \centering
    \includegraphics[width=.37\textwidth]{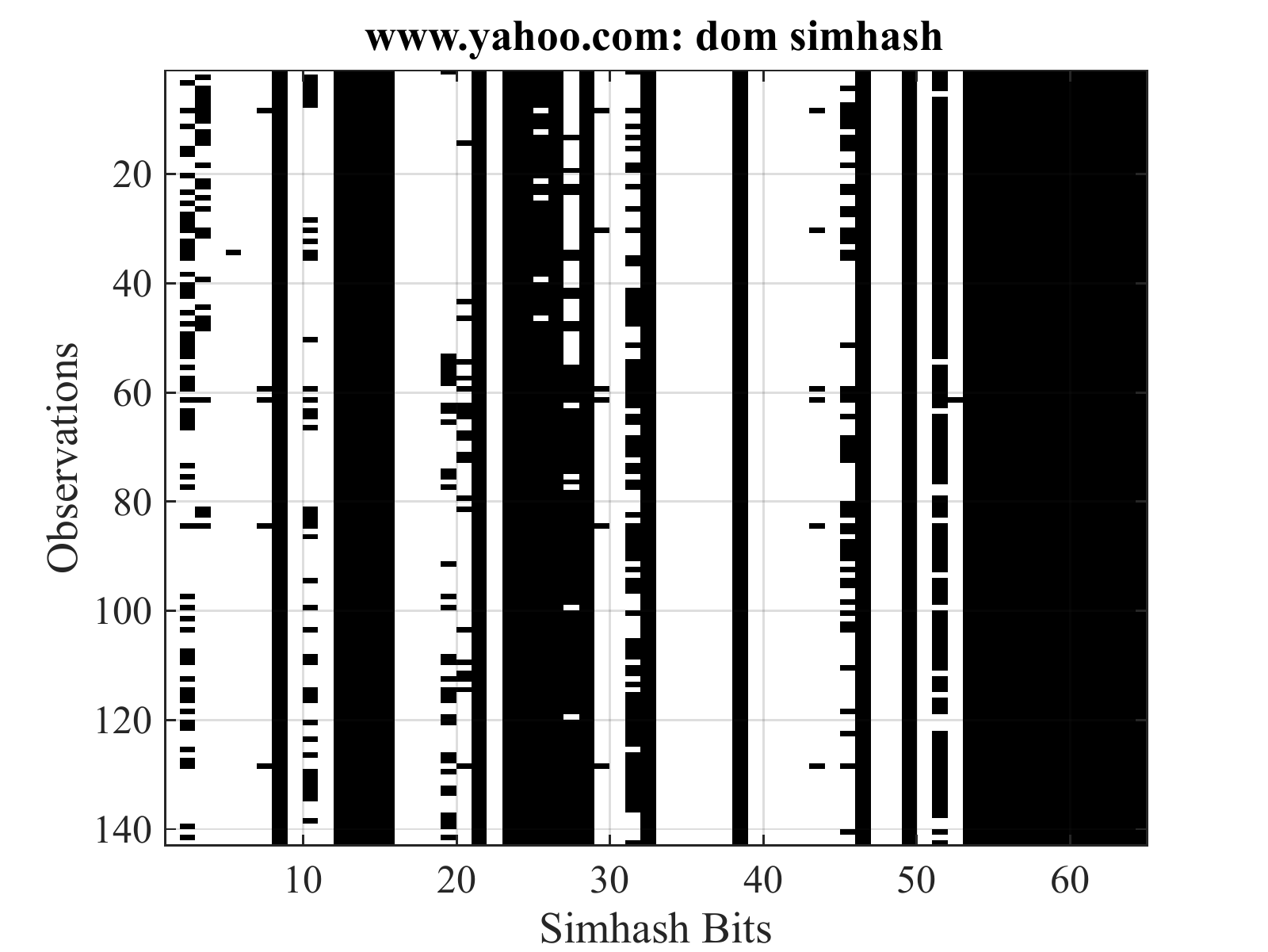}

    \label{fig:yahoo-dom-user}
  }
	\vspace{-0.5em}      
  \caption{Yahoo Simhash changes over 7x24 period Feb.1-7, 2015}
 	\vspace{-1.5em} 
  \label{fig:yahoo-simhash}
\end{figure*}

~\autoref{fig:yahoo-simhash} gives an example on how Simhash values for text and tag features changes along time. These graphs show changes of \textit{www.yahoo.com} over 7x24 period from
Feb.1, 2015 to Feb.7, 2015. ~\autoref{fig:yahoo-text-user} shows text Simhash
and ~\autoref{fig:yahoo-dom-user} shows tag simhash. The x-axis is bits of the Simhash value and
y-axis is the id of each observation (id increases in the order of collection time).
According to ~\autoref{fig:yahoo-simhash}, text Simhash changes rapidly, indicating dynamic nature of yahoo, and tag Simhash changes less and slower.

\subsubsection{Clustering}
\label{sss:clustering}
Websites can have multiple versions of content for many reasons.
An example is the change (expiration) of domain ownership, where a domain returns completely content all of a sudden. Since we are monitoring websites over a period of time, we need to be able to
build models for different versions of websites and learn patterns for each version.

We use agglomerative hierarchical clustering ~\cite{jones2014scipy} to cluster
the collected Simhash values. It is a bottom up approach: each observation starts in its
own cluster, and pairs of clusters are merged as one moves up the hierarchy.
In order to decide which clusters should be combined, a measure of dissimilarity
between sets of observations is required. In most methods of hierarchical
clustering, this is achieved by use of an appropriate metric (a measure of
distance between pairs of observations), and a linkage criterion which specifies
the dissimilarity of sets as a function of the pairwise distances of
observations in the sets. This work represents each Simhash as a 64-bit vector and specifies hamming distance as distance metric. The linkage method used is the average distance and the linkage criterion is inconsistent coefficient.

\textbf{Average linkage.} 
Let $S_{spider} = \{s_{i}, i \in (1,n)\}$ be the Simhash values of $n$ spider views for a website, and 
$s_{user}$ be the user view for this website. We use the distance from $s_{user}$ to centroid of $S_{spider}$ as a measure of similarity between $s_{user}$ and $S_{spider}$.
Consider Simhash values as 64-dimension vectors, we can compute the centroid $S_{spider}$ by summing each dimension of Simhash in the cluster and divide it by the cluster size.

\textbf{Inconsistent coefficient.} This coefficient characterizes each link in a cluster tree by
comparing its height with the average height of neighboring links at the same
level and below it in the tree. The higher the value of this coefficient, the less similar the
objects connected by the link. By using threshold of inconsistency
coefficient as criterion, we could get several clusters for each website.

~\autoref{coefficient:define} explains how inconsistent coefficient is computed.
$\alpha$ is inconsistent coefficient, $d$ is the distance between two
clusters, $\mu$ is mean of the heights of all the links included in the
calculation, $\sigma$ is standard deviation of these heights.
Each calculation in our system includes all links on the same level and below it.
\begin{equation}
  \label{coefficient:define}
  \alpha  = \frac{d - \mu}{\sigma}
\end{equation}

Let $T_{learn}$ denote the threshold for inconsistent coefficient, ~\autoref{coefficient:learn} shows the merging criterion in clustering phase. For leaf nodes, nodes that have no further nodes under them, the coefficient is set to 0.
\begin{equation}
\label{coefficient:learn}
\begin{gathered}
S_{spider_{new}} =  S_{spider, r} \cup S_{spider, s} \text{ if } \alpha < T_{learn} \\
\text{where }
\alpha = \frac{d_{r,s} - \mu}{\sigma}, \\
d_{r, s} = d(centroid_r, centroid_s)\\
\mu = avg(x), x \in  links_r \cup links_s, \\
\sigma = std(x), x \in links_r \cup links_s \\
S_{spider, r} = \{s_{r, i}, i \in (1, n_r), centroid_r, links_r\}, \\
S_{spider, s} = \{s_{s, j}, j \in (1, n_s), centroid_s, links_s\}, \\
\end{gathered}
\end{equation}

After hierarchical clustering, $S_{spider}$ is divided into $c$ clusters $S_{spider, k}, k \in (1, c)$.
We denote each cluster $S_{spider, k}$ with the centroid and links used in
clustering phase $S_{spider, k} = \{centroid_k, links_k\}$. This representation is
intended for comparison with new observations (single node clusters). $centroid$ is used to
compute distance from new observations to existing clusters, $links$ are used
to compute $\mu$ and $\sigma$.
These compact representations are the Simhash-based Website Models for this website.
By limiting the maximum number of valid spider copies, SWM can represent websites in $\mathcal{O}(1)$ space.

\subsection{Cloaking Detection}
For a particular website, we compare what users see on the client side with models fetched from the server to find cloaking, i.e. compare $s_{user}$ with $S_{spider, k}, k \in (1,c) $ to find outliers.
In the clustering phase, we use inconsistent coefficient to test similarity between clusters and decide whether we should merge two clusters. Since higher inconsistent coefficient $\alpha$ indicates dissimilarity, we use the this coefficient to test outliers as well. Let $T_{detect}$ denote the threshold used to test outliers. Consider the user view $s_{user}$ as a single node cluster $\{s_{user}\}$, and the outlier test is to check whether $\{s_{user}\}$ is similar to any of $S_{spider, k}, k \in (1,c) $. If $\alpha_k = (d_k-u_k) / \sigma_k > T_{detect}, \forall S_{spider, k}$, we consider $s_{user}$ as an outlier, i.e. cloaking candidate.

However, in reality, there can be consistent differences between what spiders
see and what users see. For example, a website shows the ad-free version to spiders, but add ads to the user version.
Moreover, there are websites that rarely changes, meaning standard deviation $\sigma$ is
zero and coefficient $\alpha$ is undefined. A fixed $T_{detect}$ cannot handle such case.
To tolerate errors introduced by the two problems, we introduce a minimum radius $R$. The modified formula for detection is ~\autoref{radius:detect}:
\begin{equation}
  \label{radius:detect}
  \text{If } d_k - R - \mu_k > T_{detect} \sigma_k, \forall S_{spider, k}, \text{reject } s_{user}
\end{equation}
~\autoref{s:impl} implements the proposed SWM and cloaking detection
algorithm, and provides insights and suggestions on the selection of $T_{learn}, T_{detect},
R$.

\subsection{Cloaker Catcher}
\label{ss:cloaker-catcher}

\begin{figure}[h]
	\centering
	\includegraphics[width=.45\textwidth]{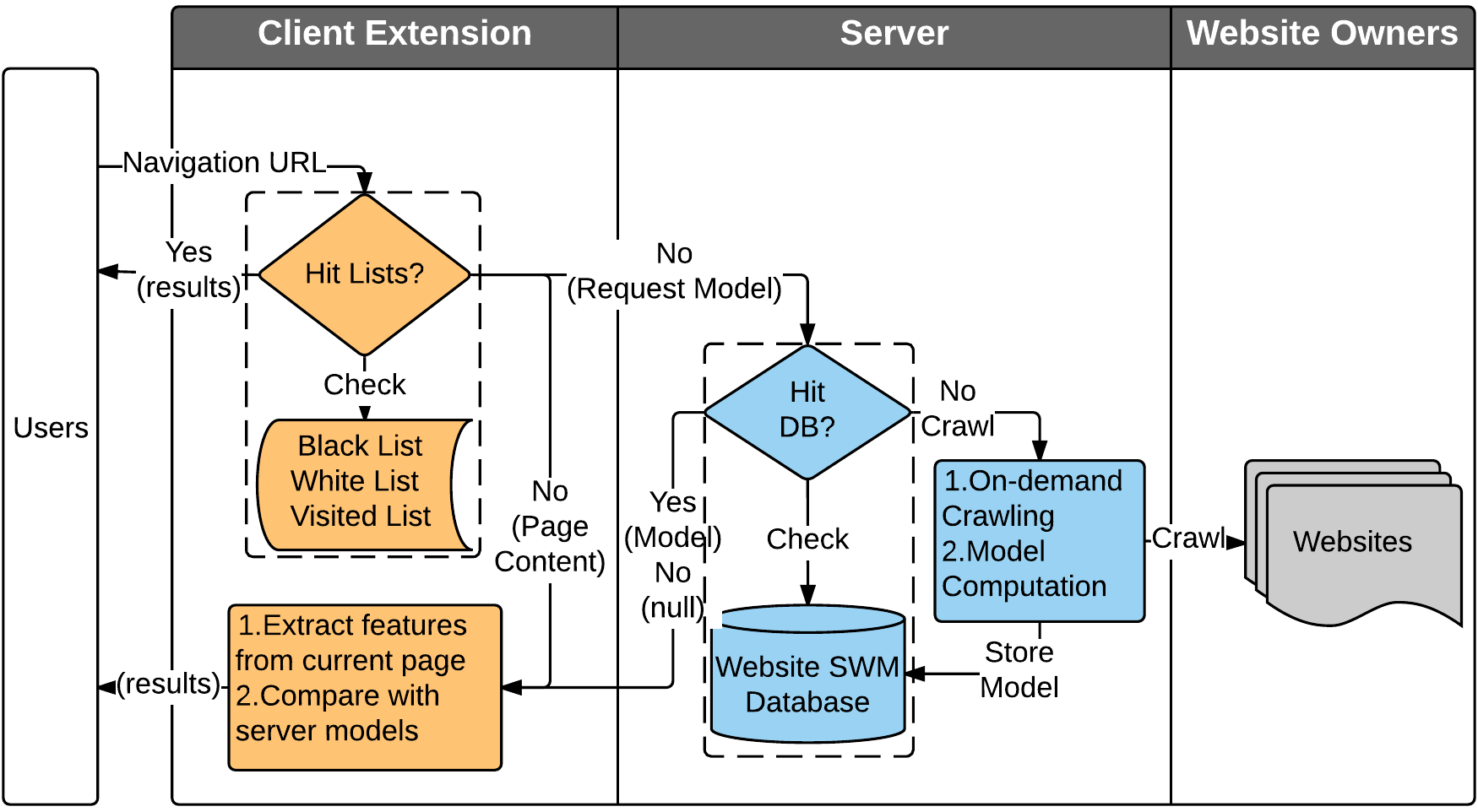}
    \vspace{-1em}
	\caption{Workflow of Cloaker Catcher}
    \vspace{-1.5em}
	\label{fig:workflow}
\end{figure}
Based on SWM and the cloaking detection algorithm, we propose Cloaker Catcher, a system consists of server side crawling and client side detection to combat cloaking.
~\autoref{fig:workflow} shows the workflow of Cloaker Catcher. We implement the client side as a chrome extension, but it is general to all modern browsers that supports extension. The server side is a program that crawl websites and store SWMs in database. When the client requests for a specific URL, server lookup the database and send the SWM to the client. The client side then compares local page content with server models to find cloaking.

In order to be accurate and efficient, we add several mechanisms and modules in the system. (A) On the client side, cloaking queries are triggered only when users click through search engine results and ads (currently configured to trigger on Google search). (B) The navigation URL landed from search engines are checked against three lists, i.e. a black list, a white list and a visited list, and is classified immediately if contained in any of the three.
The black list and the white list contain URLs that are known bad or good. The two lists are commonly used in other client-based detection literature ~\cite{rajab2013camp}. The visited list is unique to our system and it contains the websites that users have visited before. The visited list is considered good. This is based on the intuition that, cloakers will abuse users on the first visit. If a scammer want to make a sale, he has to show scam content related to the searched keyword on the first visit, otherwise, users may find this website irrelevant and leave forever.
With the help of visited list, we can reduce number of cloaking queries and reduce false positives from websites that customize content for individual users. For instance, websites personalize content for logged in users (tracked through cookies). We collect the visited URL list by checking the browsing history and cookies in users' browser.
(C) If the URL is not in any of the three lists, the extension sends the URL to the server and gets SWMs back. The extension computes Simhash values for current document and compares it with server side SWMs.
(D) On the server side, we maintain a SWM database of the crawled websites and reply to users' request with the corresponding SWM. If the requested URL is not in the database, the server first returns immediately and start the on-demand crawling module, which crawls the websites every M hours for N times (currently, M = 1, N = 5). The on-demand crawling module can be configured to crawl from Google IPs. This is done by querying Google Translate~\cite{wang2011cloak}.

\section{Implementation}
\label{s:impl}
There are three parameters to be learned in Cloaker Catcher, the upper bound of inconsistent coefficient in the clustering phase $T_{learn}$, the lower bound of inconsistent coefficient  $T_{detect}$ and the minimum radius $R$ in the detection phase.
In this section, we describe how we collect four different datasets, two for SEO and the other two for SEM, and how we get the groundtruth. The groundtruth is used to train and select parameters. The four different datasets are used to measure current cloaking state in SEO and SEM in ~\autoref{s:measurement}.

\subsection{Dataset}
\label{ss:dataset}
Popular words in SEO and SEM are different. Because high search volume doesn't directly imply high ad value. Therefore, we collect keywords and landing pages for SEO and SEM separately.

\subsubsection{Keywords}
\textbf{SEO.} Similar to Wang et al. ~\cite{wang2011cloak}, in order to detect and measure cloaking that intended to gather high volumes of undifferentiated traffic, and those target on specific cloak search terms, we collect two set of words, hot search words $W_{hot, search}$ and abuse
oriented words $W_{spam, search}$. $W_{hot, search}$ consists of 170
unique monthly hot search words
from Google trend ~\cite{google-trend} from Jan 2013 to Dec 2014. $W_{spam,
search}$ is first manually collected by referring to Google's search and ad policy for
basic abuse oriented words in search engine. The abuse categories include 
gaming ad network, adult-oriented content, alcoholic beverages, dangerous products, dishonest behavior, gambling-related content, healthcare and medicines. We then expand it using Google Suggest, and get 1024 words as $W_{spam, search}$ .

\textbf{SEM.} Popular search words doesn't imply high ad value. For example, navigational keywords, such as facebook, have low ad value, because users just want to navigate to \textit{facebook.com} instead of going anywhere else or buying anything. Therefore, we collect monetizable words for SEM related data collection. We measure monetizability of words by querying Google Keyword Planner (GKP) ~\cite{keyword-planner}. GKP provides convenient API for checking competition
\footnote{Competition is a [0,1] scale measure, 0 means nobody bids for it and 1 indicates high bidding competition.} 
and suggests bid price for keywords. In SEM, we collect two sets of keywords, trending and abuse-oriented ones.
The former is obtained by using GKP to filter a large set of hot keywords, i.e. 31679, from Google Trend~\cite{google-trend} and select words that have more than 1k monthly search, greater than 0.1 competition and higher than 0.1\$ suggested bid price. As a result, we get 11622 keywords as $W_{hot, ad}$. The abuse-oriented set, denoted by $W_{spam, ad}$, is obtained by removing words that has no competition and no bid price in $W_{spam, search}$.  $W_{spam, ad}$ ends up with 573 words.


\subsubsection{Crawling}
\label{sss:crawling}
Starting from the four collected keyword sets, we automate browsers to do
search-and-click tasks. This work uses Selenium~\cite{selenium}, an open source
browser automation tool, to visit websites while mimicked as users and spiders.
We use Selenium to automate a JS-supported browser for two reasons: detect Javascript based cloaking and increase coverage \footnote{In our early experiments, around 20\%
of advertisements have Javascript redirection and cannot be retrieved by tools with no JS support, such as wget and curl.}

When dealing with SEO keywords, we set browser user-agent as \\
\centerline{\scalebox{.8}{``Googlebot/2.1 (+http://www.google.com/bot.html)"}} \\
to mimic Google bot and \\
\centerline{\scalebox{.8}{``Mozilla/5.0 (Windows NT 6.3; Win64; x64) AppleWebKit/537.36 }}\\
\centerline{\scalebox{.8}{(KHTML, like Gecko) Chrome/37.0.2049.0 Safari/537.36"}} \\
to mimic Chrome users on Windows machine.
For SEM keywords, settings for users are the same, but the bot user-agent is set as \\ \centerline{\scalebox{.8}{``AdsBot-Google (+http://www.google.com/adsbot.html)"}} \\
to mimic Google ads bot. AdsBot is used by Google to inspect ads, and it is required by Adwords Policy~\cite{google-ad-policy} that landing pages should show consistent content to AdsBot and normal users.

In SEO and SEM, we use multiple IPs to visit and crawl websites because we get blocked frequently by Google and changing IP enables us to continuously do search-and-visits. Similar to previous works, we don't attempt to get Google IP or real users in data collection and we might miss cloakers that employs IP cloaking in our measurement. However, this doesn't affect our model training because it is trained generally for differentiating dynamic websites against cloaking ones.
Our crawling process is:
(A) For each word in $W_{hot, search}$, $W_{spam, search}$, $W_{hot, ad}$, $W_{spam, ad}$, we search in Google and click on results in first 20 pages (top 200) or ads in first 5 pages as normal user. We save landing URLs $URL_{landing, user}$ and web pages to disk.
(B) For landing URLs collected in step (A), we directly visit them 6 times
(because we need multiple spider copies to learn clusters) as Googlebot or AdsBot and
save website content to disk. This gives us separate dataset for the four keyword sets, denoted as 
$D_{spam, search}$, $D_{hot, serch}$, $D_{spam, ad}$ and $D_{hot, ad}$.

\subsubsection{Preprocessing}
The obtained four data sets in the crawling phase. Since we want to model website on a per URL basis and parameters in URL may change for every visit, it is necessary to define the granularity of
comparison, i.e. what is a unique URL. For example, ad campaign information are encoded as parameters in GET requests, and is different every time, making it difficult to compare based on full URL. Therefore, we preprocess the URLs by (1) Strip URL parameter values (2) Keep parameter names, (3) Discard scheme (4) Remove fragments. For instance,
\textit{http://www.example.com/?user=value\#fragment} is simplified to
\textit{www.example.com/?user}. 
The resulting strings are used as unique identifiers for URLs.
After preprocessing, $D_{spam, search}$ consists of 129393 unique URLs, $D_{hot, search}$ has 25533, $D_{spam, ad}$ has 2219, and $D_{hot, ad}$ has 25209.
\footnote{Automating website visits through selenium
and chrome sometimes results in incomplete rendering or empty content, these
examples are removed from collected dataset}.

\subsection{Groundtruth}
We collect groundtruth semi-automatically. Starting from $D_{hot, search}$ and $D_{spam, search}$,
we first remove duplicates (same signature between user views and Google views), because they are not helpful for training of SWM which is designed to handle dynamics of websites.
Next, we randomly sample and manually label remained websites from $D_{hot, search}$ and $D_{spam, search}$ using heuristics such as domain reputation~\cite{wot}. 
It is important to notice that, although the labeling process uses URL
reputation information (highly reputed domains are less likely to do cloaking),
it is orthogonal to the algorithm's capability, because the algorithm only measures difference of text and tag features between the original documents. Through the massive labeling process, we flagged 1195 cloaking examples. In terms of normal websites, we randomly select 5308 samples from the non-cloaking dataset. The two parts, 6503 URLs in
total, are combined as groundtruth $D_{g}$ for training and evaluation.
The websites in $D_{g}$ are converted to Simhash values $S_{g, text}$ and $S_{g, tag}$ using the feature extraction and dimension reduction technique (Simhash) described in ~\autoref{ss:swm}.

\subsection{Model Training}
\begin{figure}[h]
	\centering
	\includegraphics[width=.45\textwidth]{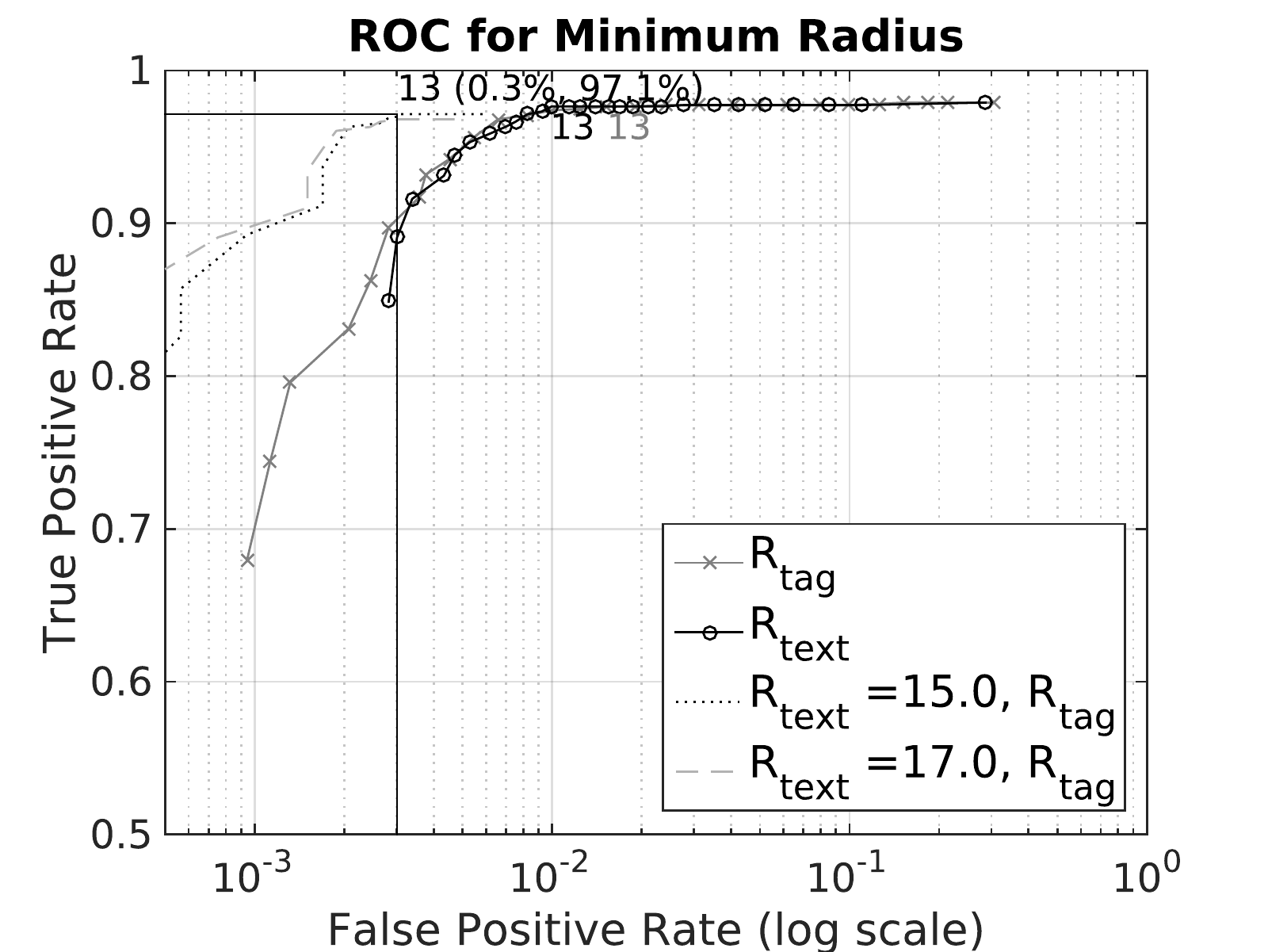}
	\vspace{-1em}
	\caption{ROC for detection based on individual features and combination of them}
	\vspace{-1em}
	\label{fig:roc}
\end{figure}

\begin{table*}[t] 
	\renewcommand{\arraystretch}{1.3}      
	\caption{Cloaking Distribution}      
	\label{tbl:result}
	\centering
	\begin{tabular}{c|c|c|c|c|c|c|c|c|c|c|c}
		\hline
		\multirow{2}{*}{ Category } & \multicolumn{4}{c|}{Traffic Sale} &
		\multirow{2}{*}{PPC} & \multirow{2}{*}{Error} & \multirow{2}{*}{IS} &
		\multirow{2}{*}{Phishing} &
		\multirow{2}{*}{PD} &  \multirow{2}{*}{Malware} & \multirow{2}{*}{Total}\\
		\cline{2-5}
		& Pharmacy & Gamble & Loan & TS  & & & & & & &\\
		\hline
		Spammy Search & 26.5\% & 60.8\% & 1.3\% & 1.1\% & 1.1\% & 1.7\% & 4.9\% & 0.7\% & 2.9\% & 0.8\% & 2491 \\
		Hot Search & 35.5\% & 2.2\% & 30.1\% & 29.0\% & 0 &  2.2\% & 2.2\% & 0 &  3.2\% & 0 & 93\\
		Spammy Ads & 0 & 0 & 0 & 0 & 14\% & 0 & 86\% & 0 & 0 & 0 & 7\\
		Hot Ads & 0 & 0 & 0 & 35\% & 0 &  0 & 65\% & 0 & 0 & 0 & 17\\
		\bottomrule
		\multicolumn{12}{c} {TS: Other Traffic Sale, PPC: Pay-Per-Click, IS: Illegal
			Service, PD: Parking Domain}
	\end{tabular}
	\vspace{-1.5em}
\end{table*}

$T_{learn}$ and $T_{detect}$ are used to generate an adaptive threshold for each website and $R$ is to make system robust to consistent difference between spider and user copies.
Based on labeled dataset $D_{g}$, our target is to learn the three parameters. Since $R$ is parameter unrelated to page dynamics, we first select optimal $T_{learn}$ and $T_{detect}$, and then evaluate our algorithm for different settings of $R$.

Because $R$ is a parameter to allow the system to handle consistent difference between spider and user copies, we first set $R$ to be zero, and do a five-fold stratified cross validation using Scikit-learn~\cite{scikit-learn} on $D_{g}$. In the learning phase, our objective function is to minimize the total number of classification errors. If the total numbers are the same, we select the one with minimal $d = T_{detect} - T_{learn}$ since $d$ represents the range of coefficients that are neither similar nor dissimilar. After the five-fold stratified cross validation process using the described objective function on $S_{g, tag}$ and $S_{g, text}$  for optimal parameter selection, we find the optimal parameters:
$T_{detect, tag} = 1.8$, $T_{learn, tag} = 0.7$, 
$T_{detect, text} = 2.1$ and $T_{learn, text} = 0.7$.

Similarly, $R_{text}$ and $R_{tag}$ are decided separately. In this section, we conduct three experiments using different features: (1) tag only (2) text only (3) combination of text and tag features. We apply five-fold stratified cross validation and the same objective function to learn and test $S_{g, tag}$ and $S_{g, text}$. The optimal parameter for tag Simhash is $R_{tag} = 17$, text Simhash is $R_{text} = 16$. Cross and oval marked lines in ~\autoref{fig:roc} shows the ROC curve for $R_{tag}$ and $R_{text}$. Next, in order to show the combined performance for tag and text features, we set $R_{text}$ to 15 and 17 separately and change $R_{dom}$ as shown in dotted lines in ~\autoref{fig:roc}.

It is straightforward that combining tag and text signature improves performance. 
The learned parameters are used in ~\autoref{s:measurement} to detect and measure cloaking
on the four datasets, and is set to $R_{text} = 15$ and $R_{tag} = 13$,  $T_{detect, tag} = 1.8$, $T_{learn, tag} = 0.7$,  $T_{detect, text} = 2.1$ and $T_{learn, text} = 0.7$, which corresponds to 0.3\% FPR and 97.1\% TPR. 
It is important to notice that, 0.3\% FPR is the result on our reduced dataset, which is obtained by removing websites that have doesn't change content or structure (tags) based on visitors. If we consider the whole dataset, our false positive rate is much lower than 0.3\%FPR.
The selected parameters are also used to configure Cloaker Catcher and preliminary feedbacks from several users show that our system is effective in detecting both SEO and SEM cloaking.
\section{Measurement}
\label{s:measurement}
\subsection{Terminologies}
We detected and measured cloaking in
four collected datasets: spammy search, $D_{spam, search}$, hot search,
$D_{hot, search}$, spammy ads, $D_{spam, ad}$, hot ads, $D_{hot, ad}$. 
 Based on summarization of SEO cloaking types in ~\cite{wang2011cloak} and our observations of SEM cloaking, we categorized cloaking into seven types:
Traffic Sale, Pay-Per-Click (PPC), Error Page, Illegal Service, Phishing, Parking Domain, and Malware.
To better analyze cloaking incentives, we further divided traffic sale into four categories: pharmacy, gambling, loan, and general traffic sale.  Traffic sale sites are usually from third parties and include iframe pointing target sites for traffic monetization. PPC means that landing page host pay-per-click advertisements.
Error Page refers to websites that deliver erroneous information. Illegal Service includes websites that provide illicit services such as drugs, essay writing, copyrighted content, and bot service. Parking Domain are domains that redirect users to unwanted download, but not necessarily malicious.

\subsection{Sumamry of Findings}
We measured the cloaking state of hot and abuse-oriented words in the four collected datasets separately. After manually inspection of websites related to abuse-oriented words, we found that 2\% of the search results and 0.32\% of the ads are cloaking. For hot words, 0.36\% of the search results and 0.07\% of the ads are cloaking. An interesting finding is that the SEO cloaking sites mainly targets traffic sale, while the SEM cloaking websites provide illegal services. 

In $D_{spam, search}$, we apply our cloaking detection system on 129393 websites. We manually label the detected results and identify 2491 cloaking websites. We further analyzed incentives of cloaking websites and show the results in ~\autoref{tbl:result}. The majority of these websites fall into traffic sale. In total, websites related to phishing, parking domain and malware sums to 110, a non-negligible amount missed by Google because of cloaking. In $D_{hot, search}$,  we found 93 cloaking websites.
Among these sites, pharmacy, loan and general traffic sale are prevalent categories.
After combining observations of $D_{spam, search}$ and $D_{hot, search}$, we concluded that the main goal of SEO cloaking is to achieve traffic sale.
In spammy ads $D_{spam, ad}$, we processed 2219 websites and identified 7 cloaking websites. 6 of them provides illegal service. In $D_{hot, ad}$, we checked 25209 websites and found 17 cloaking websites. 11 of them provides promoting illegal services and 6 provides traffic sale.

\section{Performance}
\label{s:performance}
Generally, there are mainly two phases in cloaking detection literatures: feature extraction and comparison. In the feature extraction phase, these algorithms traverse the web pages and extract statistical and semantic features. In the comparison phase, the algorithms compare the features extracted from both user copies and spider copies to identify cloaking.
In this section, we present a taxonomy of cloaking detection methods then explain differences between Cloaker Catcher and other cloaking detection works. 
Since our goal is to design a client-based detection system in which performance matters, we re-design previous works as client-based systems similar to Cloaker Catcher and compare time complexity and storage of our system with them to show that our system is a better fit for client-based cloaking detection. Secondly, we measure the computation and comparison performance using real world examples and show that the average delay is around 100 milliseconds.

\subsection{Efficiency Comparison}
We provide a taxonomy of cloaking detection methods based on Lin's summary in his tag-based system ~\cite{lin2009detection}. ~\autoref{comparison} shows the comparison between Cloaker Catcher and previous systems along six dimensions, features, max f1-score, IP/SEM cloaking, feature extraction time, amount of data transmitted, and comparison time.

\begin{table*}[t]
	\renewcommand{\arraystretch}{1}
	\caption{Comparison of cloaking detection methods}
	\label{comparison}
	\centering
	\begin{tabular}{c|c|c|c|c|c|c}
		\hline
			Methods & Features & Max F1 Score & IP/SEM & Extraction Time & Data Transmitted & Comparison Time\\
			\hline
			Najork ~\cite{najork2003system} & W, L, T & low &  \cmark & $O(N_W + N_L + N_T)$ & $O(1)$ & $O(1)$ \\
			Term \& Link Diff ~\cite{wu2005cloaking} & W, L & medium  & \xmark   & $O(N_W +N_L) $ & $O(N_W+N_L) $ & $O(N_W+N_L) $\\
			Wu \& Davison ~\cite{wu2006detecting} & W, L & medium & \xmark  & $O(N_W +N_L) $ & $O(N_W+N_L) $ & $O(N_W+N_L) $\\
			CloakingScore ~\cite{chellapilla2006improving} & W & medium & \xmark   & $O(N_W)$ & $O(N_W)$ & $O(N_W)$ \\
			TagDiff ~\cite{lin2009detection} & W, T & high & \xmark  & $O(N_W + N_T)$ & $O(N_W + N_T)$ & $O(N_W + N_T)$ \\
			Dagger ~\cite{wang2011cloak} & W, T & high & \xmark & $O(N_W+N_T)$ & $O(N_T)$ & $O(N_T)$ \\
			Hybrid Detection ~\cite{deng2013uncovering} & W, L, T & high & \xmark   & $O(N_W+N_L+N_T)$ & $O(N_L+N_T)$ & $O(N_L+N_{T}^{2})$\\
			Cloaker Catcher & W, L, T & high & \cmark & $O(N_W + N_T)$ & $O(1)$ & $O(1)$ \\
			\bottomrule
			\multicolumn{7}{c} {$W$: words, $T$: tags, $L$: links}
	\end{tabular}
	\vspace{-1.5em}
\end{table*}

The ``features" dimension specifies what features from a web page are used for cloaking detection. 
Possible features are terms and tags, because an HTML web page comprises these three types of elements (Javascript, CSS etc. are included files). These elements are usually used in combination to reduce false positive.

The ``max F1-score" dimension is summarized empirically based on comparison presented in ~\cite{lin2009detection} and ~\cite{deng2013uncovering}. Since the two works compare the  precision and recall of these algorithms, we use F1-score, i.e. $F_1 = 2\cdot precision\cdot recall / (precision + recall)$ to summarize them. Generally, the higher the F1-score, the better the system. 
Najork's work ~\cite{najork2003system} uses cryptographic hash functions to fingerprint a website and have high false positive because they can not handle dynamic website. 
Relative F1 scores for ~\cite{wu2005cloaking,wu2006detecting, chellapilla2006improving, lin2009detection} are directly inferred from ~\cite{lin2009detection}. TagDiff~\cite{lin2009detection}, Dagger~\cite{ wang2011cloak} and our system, Cloaker Catcher, use the same types of features, and achieves similar F1-score. Hybrid Detection~\cite{deng2013uncovering} adds links to the feature set, and achieves slightly better results. Although adding links may improve the performance of our system, we skip it for now because on one hand links are not suitable for Simhash algorithm, and on the other hand directly comparing link sets introduces too much overhead.

The ``IP/SEM" dimension specifies whether the original system is capable of detecting IP cloaking and SEM cloaking. Cloaker Catcher and Najork's system ~\cite{najork2003system} are the only systems capable of detecting them since they are client-based. However, the poor precision of the latter prevents its usage.

The ``Extraction Time", ``Data Transmitted" and ``Comparison Time" columns are not trivial because some of the original systems are not designed for client-based detection.
To make these systems comparable to our system, we modify the previous works in a client-server way: (1) The server caches the features for a large amount of websites and responds to client requests with features for the requested URL (2) The clients request features for the current URL from the server, extract features from the client copy and compare the two sets of features to identify cloaking. This design guarantees that the server could get no more information than the URLs, which we can further use techniques employed in existing APIs such as SafeBrowsing API~\cite{safebrowsing} to reduce user linkability. The time and data requirements are shown using number of terms, tags and links.
As shown in the ``Data Transmitted" column, Cloaker Catcher benefits from compact website models and transfers much less data compared to Dagger~\cite{wang2011cloak} and Hybrid Detection ~\cite{deng2013uncovering}. This is important in client-based detection because network delay is unpredictable. In terms of ``Extraction Time" and ``Comparison Time", the most similar system, Dagger uses text shingling to reduce number of term comparison, but they still do pairwise comparison of tags. In order to detect cloaking on the client side, their clients need to traverse the document in $O(N_W + N_T)$ time to extract features, request  $O(N_T)$ features including fixed number of shingles and all the tags from the server and compare these features in $O(N_T)$. Cloaker Catcher, in contrast, uses fixed size features and compares much faster.

\subsection{Performance of Cloaker Catcher}
\begin{figure}[h]
	\centering	
	\includegraphics[width=.5\textwidth]{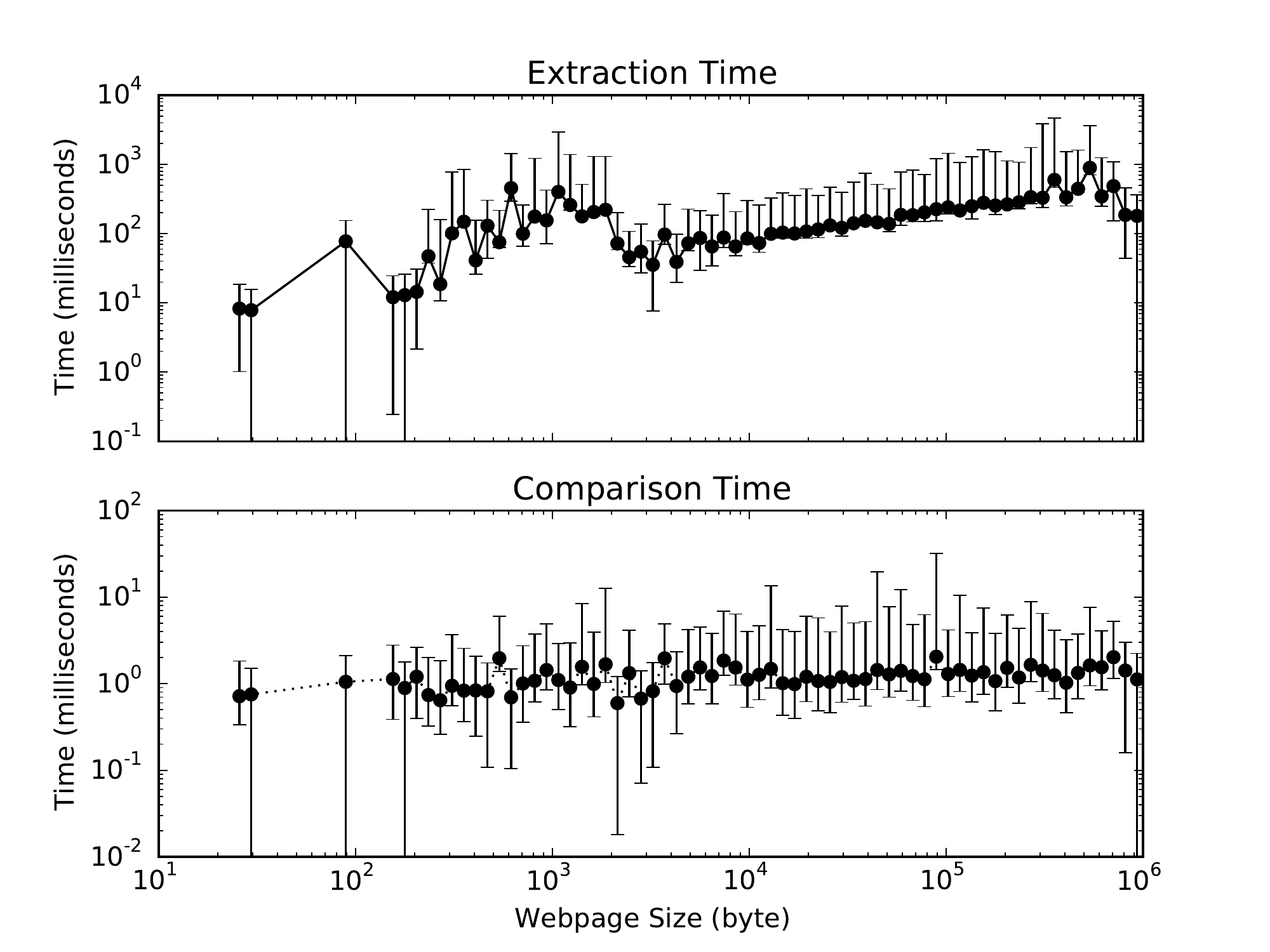}
	\vspace{-2em}	
	\caption{Extraction and Comparison Overhead introduced by Cloaker Catcher}
	\vspace{-1em}	
	\label{fig:performance}
\end{figure}

As described in ~\autoref{fig:workflow}, Cloaker Catcher detects cloaking in several steps, in order to minimize number of server queries and detection overhead.
First, it only examines the websites that are visited through Google search (this step can be generalized to crawler user agent of other search operators). Secondly, these websites are checked against three lists (black/white/visited). When the websites are not in these lists, the system requests spider copies and compares with user copies. The main delay in the detection consists of three parts, feature extraction, data transfer, and comparison. Since data transfer time depends on the network status and is irrelevant to the web page size, we only measure the feature extraction and comparison overhead. We randomly select 2,000 URLs from our SWM database and use selenium to automate Chrome with Cloaker Catcher to visit these websites. We record the feature extraction time and comparison time and aggregate them by web page size. The results are shown in ~\autoref{fig:performance}, 
x-axis is the web page size in log scale and y-axis is the time cost. The graph show solid points which are average time, along with their upper and lower error bar. When the web page size are greater than 1KB, feature extraction time grows linearly and comparison remains constant. The curves fluctuate when page size is less than 1KB because of few sampled websites.
The overall delay introduced by Cloaker Catcher are hundreds of milliseconds in practice. 

\section{Limitation}
\label{s:limitation}
\textbf{URL Features.} URLs on a website show navigational features of websites, e.g. embedded links to images or videos. However, URL features are not straightforward and require further study to compactly represent them. We leave this for future work and anticipate a more robust client-based system with URL features.

\textbf{Incentive analysis automation.}
We manually classified the incentives of cloaking websites in this work. However, according to our observations, this process can be automated using machine learning approaches. For example, cloaking sites that simply embed an iframe is targeting traffic sale, and ones that redirects users to downloads are doing unwanted software download.

\textbf{SWM Availability.}
Cloaker Catcher includes a server that crawls websites on demand and builds SWM. However, this can be a huge amount of work as the number of users increases. Therefore, we advise search engines to adopt our prototype and detect cloaking at scale. For individual researchers, a trade-off is to selectively crawl websites and reply SWMs for them. For instance, researchers can disable the on-demand crawling module and crawl URLs for specific categories of words. Researchers can also filter the requested URLs based on domain reputation~\cite{wot} and only crawl the suspicious ones.

\section{Related work}
\label{s:relwk}
\textbf{Simhash.}
Simhash is a popular Locality Sensitive Hash (LSH) used in information retrieval. In the field of near duplicate detection, Minhash~\cite{broder1997syntactic} and Simhash~\cite{charikar2002similarity} are the most popular techniques. A large scale study~\cite{henzinger2006finding} has been done to evaluate the two algorithms. A great advantage of using Simhash over Minhash is that it requires relatively small-sized fingerprints. For example, 24 bytes Minhash and 64-bit Simhash achieves similar performance.
David et al.~\cite{wang2011cloak} employs shingling (Minhash) and our system use Simhash to gain a more compact representation of websites.

\textbf{Semantic Features.}
Previous works on cloaking detection focus on differentiating dynamic pages
from blatantly different content. Various ways to measure similarity of documents are proposed. 
A similar toolbar based approach is proposed by Najork et al.~\cite{najork2003system}.
The toolbar sends the signature of user perceived pages to search engines. But they cannot handle dynamic websites, and may raise high false positive. Wu et al. ~\cite{wu2006detecting} used statistics of web pages to detect cloaking and Chellapilla et al.~\cite{chellapilla2006improving} detected syntactic cloaking on popular and monetizable search terms. They showed that search terms with higher monetizability had more cloaked links than popular terms.
Referrer cloaking was first studied by ~\cite{wang2006detecting}. They found
a large number of referrer cloaking pages in their work.
Lin et al.~\cite{lin2009detection} introduced tag based methods for cloaking detection,
~\cite{wang2011cloak} extended previous efforts to measure dynamics of cloaking over five
months. They used text-based method and tag-based method to detect cloaking pages.
~\cite{deng2013uncovering} integrated previous efforts and proposed a text, link and tag based cloaking detection system. Inspired by previous works, we use texts, links and tags to model a website. We further compress the three feature sets to fixed-bits using Simhash and reduce complexity of pairwise comparison to $\mathcal{O}(1)$.

\textbf{Domain Features.}
Domain and content features are orthogonal and both contain information about a particular website. Lu et al.~\cite{lu2011surf} and John et al.~\cite{john2011deseo} have used domain-related information, e.g. redirection and domain reputation, to detect search engine poisoning.
A more thorough and long-term study of search engine poisoning is done by Leontiadis et al. ~\cite{leontiadis2014nearly}. Content-based detection and domain-based detection complement each other and may work better if combined.

\section{Conclusion}
\label{s:conclusion}
Detection of search engine spam and ad spam is a challenging research area. Cloaking has become a standard tool in the spammer's toolbox and added significant complexity for detection. Our work has identified previous cloaking works' inability in detecting IP-based cloaking and SEM cloaking and proposed a novel client-based system, Cloaker Catcher, to address these issues. The key component of Cloaker Catcher is the compact website model, SWM, which minimizes the storage for each website and enables fast comparison between user views and spider views. We further use Cloaker Catcher to detect and classify cloaking in collected search results and ads. We present the first analysis of cloaking in search ads, which shows that, the purpose of advertisement cloakers is mainly to provide illicit services, different from search cloakers who mainly want to monetize their traffic.

Given the high accuracy and low overhead of this approach, we believe Cloaker Catcher is a practical system for search engines to provide real-time protection for users. The cloaking detection can serve as a separate warning or be integrated into existing API to improve users' experience.

%
\bibliographystyle{abbrv}
\bibliography{p,sslab,conf}
\newpage
\appendix

\section{SEM Cloaking Example}
\begin{figure}[H]
  \centering
\subfloat[Search on Google: Essary Writing]{
  \includegraphics[width=.4\textwidth]{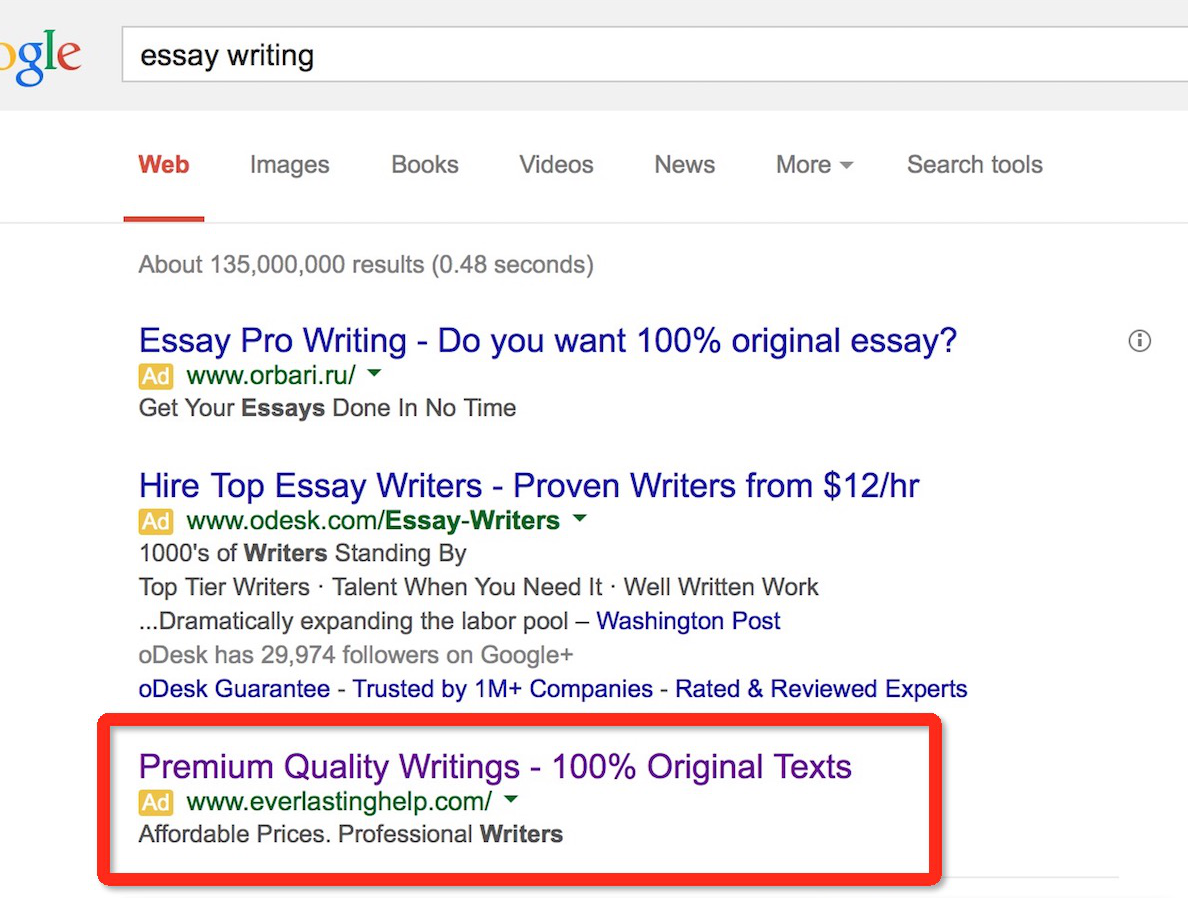}
  \label{sem:search}
}
\hfill 
\subfloat[Click-through users are presented plagiarism service]{
  \includegraphics[width=.45\textwidth]{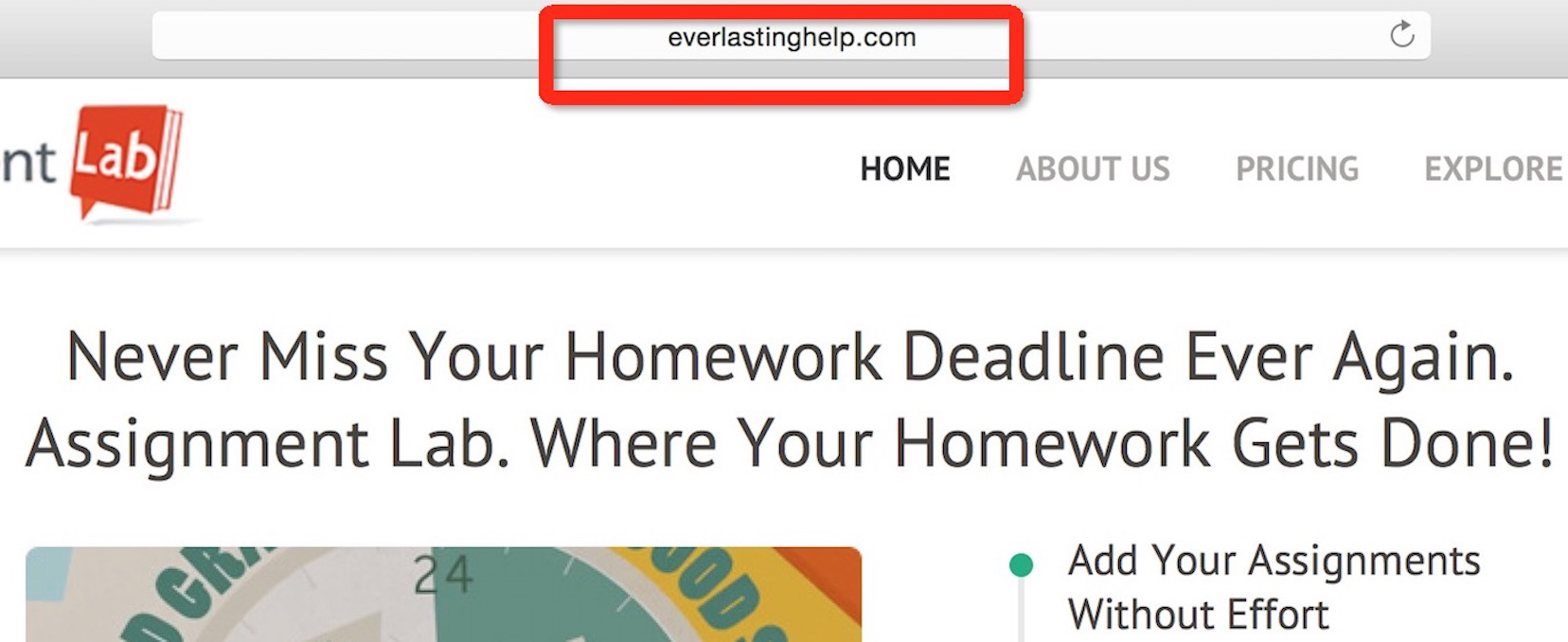}
  \label{sem:user}
}
\hfill 
\subfloat[Spiders are presented hotel page]{
  \includegraphics[width=.45\textwidth]{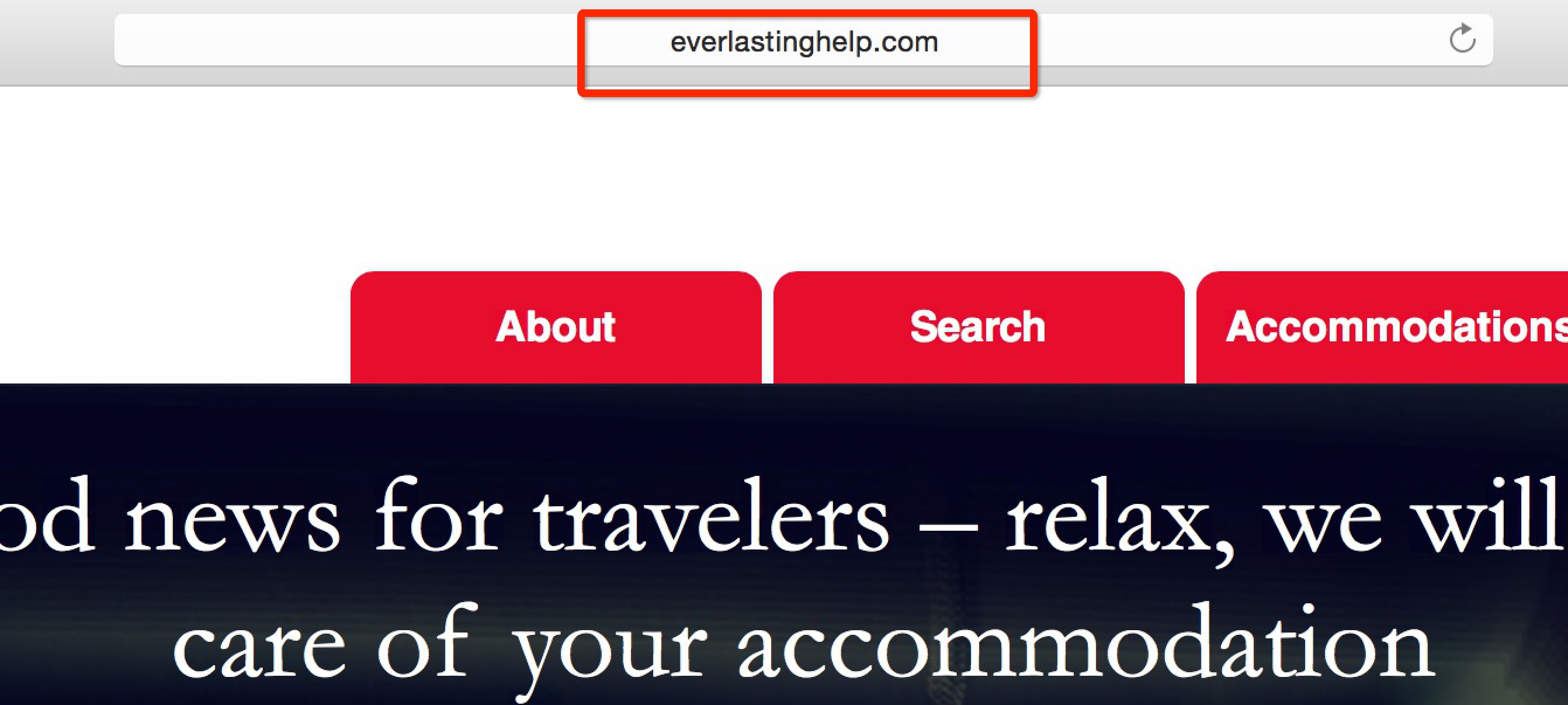}
  \label{sem:google}
}
\caption{SEM Cloaking in sponsored ads of ``essay writing"}
\label{sem:cloaking}
\end{figure}
~\autoref{sem:cloaking} shows an SEM cloaking example of keyword ``essay writing". The search result page, user view and spider view are listed.

%
%
\end{document}